# A new control system for SECRAL[*]

SU Jian-Jun(宿建军)ZHOU De-Tai(周德泰)WANG Yan-Yu(王彦瑜)ZHOU Wen-Xiong(周文雄)

Institute of Modern Physics, Chinese Academy of Sciences, Lanzhou 730000, China

**Abstract:** To improve the control efficiency of Superconducting Electron Cyclotron Resonance Ion source(SECRAL) for Heavy Ion Research Facility in Lanzhou(HIRFL), a upgrade version of control system for SECRAL is designed and set up. The control software package is developed by Visual C++, which is able to control and monitor all of the equipment for the SECRAL system about 130 parameters. The previous analog power supplies are replaced by four digital power units at High Voltage(HV) platform, The old slow speed of AC motors are upgraded to servo motors newly for higher precision, stability and linearity. Meanwhile, some control strategy and user interface are optimized. In order to prevent the incorrect operation which may cause the quench of the superconducting magnet, alarm and interlock protection functions are added to the software and hardware. Since the upgraded SECRAL control system is online, it has been running smoothly.

**PACS: 07.05.Dz, 29.85.+c          DOI: 10.11884/**

## 1.Introduction

The major HIRFL facilities is composed of the ECR ion source, fan-focused cyclotron, radioactive ion beam line, cooling storage ring, the main ring, and the experiment ring[1]. The SECRAL ion source is built to produce intense beams of highly charged ions for HIRFL. Various devices are needed to be controlled in order, for example, superconducting magnet power supplies, HV power supplies, vacuum system, beam monitor system, water temperature, pressure system, liquid helium circulation system, etc. And the picture of HV platform is shown as Fig.1.

## 2. Architecture of the hardware

Different methods are needed to control and monitor the whole systems of SECRAL for their especial working conditions, such as HV signal and the weak current of Micro-ampere level signal[2].Some equipment working in HV platform may damage the control system formed by serial bus and function module[3]. Thus, the communication between the controller and network switch is implemented through an optical-to-electrical signal converter, as shown in Fig.2. The device between the converters is a single-mode glass fiber.

Two types of special gases are required to produce a plasma: supporting-gas and working-gas. Also there are bias voltage and sputter newly. the plasma is produced through the electric discharging of the mixed gas. The quantity and flow rate of these gases must be strictly controlled and accurately monitored. The flow rate of the mixed gas is controlled by four servo motors. When the ion source is operational, a malfunction in any of the devices can result in unstable operation of the ion source, reduce the particle beam or seriously damage the devices. To protect the control devices and ensure steady operation of the ion source, interlock protection and alarm hardware is added to the remote control system.

*Supported by National Nature Science Foundation of China(u1232123)

1)E-mail:sujianjun@impcas.ac.cn

2)E-mail:yanyu@impcas.ac.cn

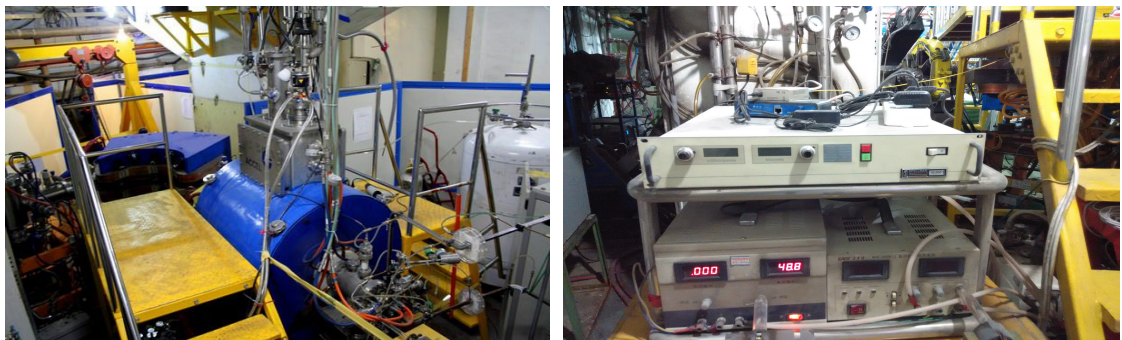

Fig.1. General view of HV platform installations

In the system, all of the equipment are controlled through a network switch which is connected to the Industrial PC. Serial servers are used to control the equipment through RS232/485. So the related equipment can be controlled through RS232/485 port directly. For example, Superconducting magnet power supplies, Helium Depth Indicator module(HDI), Helium Temperature Meter module(HTM), the meter to Measure the Weight of Helium(MWH), KEITHLEY(a meter used to measure beam current), Digital Meter(used to measure water temperature and pressure), etc.

At the same time, other equipment can be controlled through Ethernet. For instance, the TPG256(six ports Vacuum gauge), the High Voltage Power Supply, Compressor, Device side High Voltage, High Voltage Power Supply and Microwave Machine.

Before installing the system at SECRAL, laboratory tests has been done to confirm the suitability of the system. All of those equipment status information and control signals are obtained and processed via the control software running in the center computer through intranet.

## 3. Architecture of the software

The new control system for SECRAL system consists of three layers: remote monitoring layer, server control layer and field control layer. It implies either its hierarchical structure. The 3-layer control architecture is capable of extending the system functions and upgrading devices. The server control layer plays a key role in the control system. The TCP/IP protocol is adopted for the data transmission, and the database technology for the data processing and storage. The network is in charge of controlling and monitoring the parameters of the ion source to achieve a proper particle beam. It interconnects all the different types of control hardware and provides a way to communicate with any other system outside the SECRAL project scope.

The control layer receives the user commands and interprets them to control the field devices. The field devices send the real-time status of the SECRAL system to servers. The data servers receive these experimental data acquired by the data acquisition systems. These data are processed and stored into the experimental database. The system also provides data services for the users according to the query commands, for example, experimental data archiving, data processing, experimental data query, etc. The field devices connected to the system execute the network commands.

The control software are designed as independent classes, including the communication modules (TCP/IP or RS232/485), the data decoding module, the data display modules, the data storage modules and so on. Thus, new devices can be easily added. If a new device is added, it is only necessary to instantiate the classes and call the functions of the classes. Most of the functions are called by these classes. Data

Getting module is used to obtain data from the controller. It performs 3 function, including setting the appropriate parameters, sending the right command when the corresponding equipment works and gets data. Processing Data module would work immediately to decode the data according to the parameter format. After Data Getting module gets the useful parameter information, data is sent to the Display module for displaying. The Data Saving module uses a timer to sample and save the equipment data. When the time is out, the software would write the data to the data server. Once the error occurs, the module would send the Interlock Protection signal to the relevant controllers in case of the equipment damage. At the same time, the alarm would go-off, notifying the person on duty.

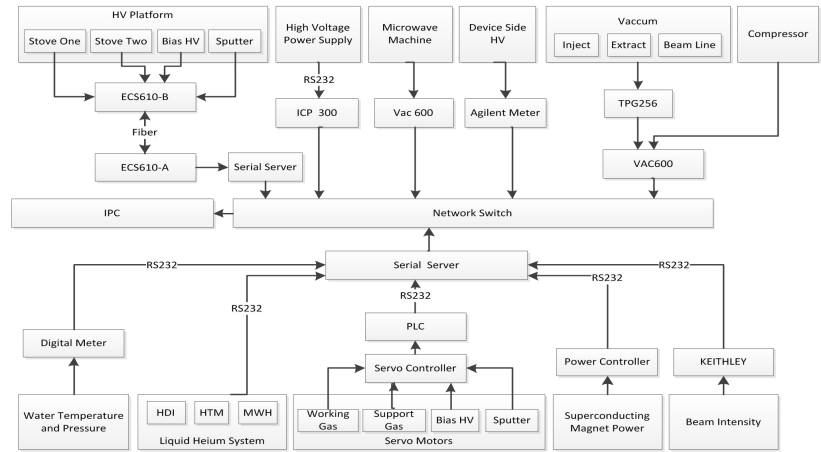

Fig.2 Schematic diagram of system.

**3.1 Control of high voltage platform**

The previous analog power supplies are discarded and four digital high powers are used at HV platform, which contain two stove power supplies, a bias voltage power supply, and a sputter power supply. The devices placed on the platform are always operating in an extremely harsh environment, in which they can be exposed to maximal voltages of 60 kV. In addition, all of the devices and controllers operate under strong electromagnetic fields, radioactive radiation and high voltage surroundings[4]. These conditions may damage the controllers. We have taken measures to solve these problems. The controller is designed with a transient voltage suppressor diode in the input port. The transient voltage suppressor can prevent damage to the controller from electrical surges. Ferrite beads are fixed around the signal wires to suppress spike signal to the controller. Thus, the controllers are protected from electrical surge damage.

**3.2 Control of servo motors**

In old SECRAL control system, two types of special gases are designed to produce a plasma, supporting-gas and working-gas. But two motors of gases were slow speed of AC motors. Now, there are four servo motors added a bias voltage motor and a sputter motor for higher control precision and linearity. The flow rate of the mixed gas is controlled with four valves driven by four accurate servo motors. To protect the controlled devices and ensure steady operation of servo motors, interlock protection and alarm function is also added to the control of four servo motors. The servo motors system is interlocked with themselves when moving distance exceed max threshold.

**3.3 design of experimental database**

A distributed experimental data processing system is developed for experimental data archiving, processing and retrieving. The data stored in the data server consist of the past experimental data and real-time display data of the device's states. The experimental data stored in the database labeled with a SECRAL shot number include engineering status (temperature, vacuum, parameters of power supplies), status of subsystems, the configuration of the devices, experimental results processed by the system status of the field devices, the information of the alarms and faults, information of the online terminals, real-time feedback values of the key parameters, the states of the control software system and so on. The system provides the data processing tools, data analysis program for analysis, statistics and visualization, and data processing and conversion tools, data query and backup program.

**3.4 Interlock and protection**

The control commands, timing sequences and logic control orders of the SECRAL system is closely related to the operational conditions of the field devices. The interlock and protection system is response for protecting operators and devices from damages by interlocking control signals with status signals. When the ion beam is extracted, the magnet power supply operation must be interlocked with the position of those components, such as beam targets and ion source isolation valves. HV platform is interlocked with the microwave machine. Once values of HV platform exceed threshold number, the microwave machine should be closed. Moreover, inject vacuum, extract vacuum and beam vacuum are interlocked with High Voltage Power Supply and Microwave Machine too. The interlock protection and alarm is implemented by a self-developed module known as the VAC600[4]. The sources used to trigger the interlock and protection alarm is implemented by "or" logic. The operators can use the control software to set the threshold for different requirements.

## 4 Results and Performance

Fig.3 show the interface of the control software, and the software has been running normally in the center control room. It is proved in practice that the control system can control and monitor all the parameters. First, reconstruction of the system guarantee running of the control system safely and stably. Second, the efficient control system reduce the frequency of maintenance for SECRAL. Finally, the upgraded system has supply stable ion beam for physics experiment for 3000 hours. Especially, the hardest $Bi^{31+}$ ion beam also has been extracted successfully. Experimental results demonstrate the upgraded SECRAL control system meets the requirements of the control functions and makes the experimental operations more automatically and visually.

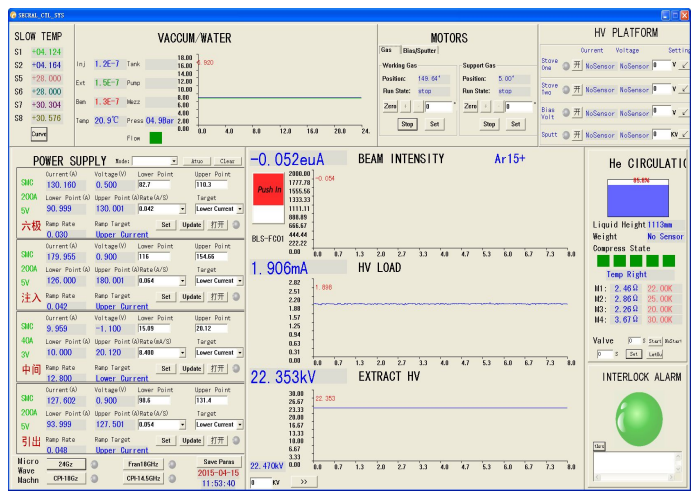

Fig. 3. Remote control software interface

Submitted to 'Chinese Physics C'[2] JiJiu Zhao. Chinese Physics C 32 (1) .(2008) :139-141

[3] Wenxiong Zhou, Yanyu Wang, et al. Nuclear Physics Review 29 (4) (2012) 364.

[4] Wenxiong Zhou, Yanyu Wang, Detai Zhou, et al. Nuclear Instruments and Methods in Physics Research A 728 (2013) 112-116.